\documentclass[twocolumn]{svjour3}
\usepackage{amsmath}
\usepackage[numbers]{natbib}
\usepackage{graphicx}
\usepackage[font=small]{caption}
\usepackage[caption=false]{subfig}
\usepackage[space]{grffile}
\usepackage{cuted}
\usepackage{float}
\usepackage{color}

\newcommand{\hs}{\hspace{0.25cm}}
\newcommand{\eqcomma}{\hs,}
\newcommand{\eqstop}{\hs.}
\newcommand{\tarea}{A^{\rm (t)}_c}
\newcommand{\farea}{A^{\rm (f)}_c}
\newcommand{\Pback}{P_{\rm back}}
\newcommand{\EB}{{\rm EB}}
\newcommand{\MSD}{\langle \vec{x}^2 \rangle}

\begin{document}

\title{Modifying continuous-time random walks to model finite-size particle diffusion in granular porous media}

\titlerunning{Modifying continuous-time random walks...}

\author{Shahar Amitai \and
            Raphael Blumenfeld}

\institute{Shahar Amitai \at
              ESE, Imperial College London, London SW7 2AZ, UK \\
              \email{s.amitai13@imperial.ac.uk}
           \and
              Raphael Blumenfeld \at
              ESE, Imperial College London, London SW7 2AZ, UK \\
              College of Science, NUDT, Changsha, Hunan, China \\
              Cavendish Laboratory, JJ Thomson Avenue, Cambridge CB3 0HE, UK}

\date{Received: date / Accepted: date}

\maketitle

\begin{abstract}
The continuous-time random walk (CTRW) model is useful for alleviating the computational burden of simulating diffusion in actual media. In principle, isotropic CTRW only requires knowledge of the step-size, $P_l$, and waiting-time, $P_t$, distributions of the random walk in the medium and it then generates presumably equivalent walks in free space, which are much faster. Here we test the usefulness of CTRW to modelling diffusion of finite-size particles in porous medium generated by loose granular packs. This is done by first simulating the diffusion process in a model porous medium of mean coordination number, which corresponds to marginal rigidity (the loosest possible structure), computing the resulting distributions $P_l$ and $P_t$ as functions of the particle size, and then using these as input for a free space CTRW. The CTRW walks are then compared to the ones simulated in the actual media.

In particular, we study the normal-to-anomalous transition of the diffusion as a function of increasing particle size. We find that, given the same $P_l$ and $P_t$ for the simulation and the CTRW, the latter predicts incorrectly the size at which the transition occurs. We show that the discrepancy is related to the dependence of the effective connectivity of the porous media on the diffusing particle size, which is not captured simply by these distributions.

We propose a correcting modification to the CTRW model -- adding anisotropy -- and show that it yields good agreement with the simulated diffusion process. We also present a method to obtain $P_l$ and $P_t$ directly from the porous sample, without having to simulate an actual diffusion process. This extends the use of CTRW, with all its advantages, to modelling diffusion processes of finite-size particles in such confined geometries.


\end{abstract}

\section{Introduction}

Diffusion plays a key role in a wide range of natural and technological processes. A textbook modelling of such processes is the consideration of the diffusion of a single memory-free particle in a given medium. The nature of such a random walk is governed by three probability density functions (PDFs): of the step size, $P_l(l_i)$; of the step direction, $P_n(\hat{n}_i)$; and of the waiting time between steps, $P_t(t_i)$. These PDFs are, in principle, position dependent, but it is standard practice to derive (or postulate) them assuming position-independence and that $P_n(\hat{n}_i)$ is uniform. The diffusion is then modelled as a continuous-time random walk (CTRW) in free space. Specifically, the CTRW is constructed by adding vectors of uniformly random orientations, whose lengths are chosen from $P_l$, at time intervals chosen from $P_t$. Averaging over sufficiently many such independent processes, the dependence of the mean square distance (MSD) on time satisfies $\MSD = Dt^\alpha$. In normal diffusion $\alpha = 1$ and $D$ is the standard diffusion coefficient. But when $P_l$ and/or $P_t$ are very wide, the diffusion might become anomalous ($\alpha \ne 1$). In particular, when $P_t$ has a slowly decaying algebraic tail and $P_l$ does not, the random walk is sub-diffusive ($\alpha < 1$) \cite{Scher1975, Scher1991}. Alternatively, if $P_l$ has a slowly decaying algebraic tail and $P_t$ does not, the random walk is super-diffusive ($\alpha > 1$), resembling a L\'evy flight \cite{Mandelbrot1983}. Diffusion processes that have the same value of $\alpha$ are said to be in the same \textit{universality class} \cite{Kadanoff1966}.

Anomalous diffusion can arise from different sources, which can only be identified by going beyond the MSD. When single particle tracking is possible, the movement can be evaluated by the time-averaged MSD (TAMSD), $\delta^2(t, T)$ \cite{Metzler2014}.While the MSD is the ensemble average of the squared distance, made during a time interval $t$,  over different realisations, the TAMSD, $\delta^2(t, T)$, is the average of the same quantity along {\it a single trajectory} of length $T$. Within the model of sub-diffusive CTRW, the TAMSD satisfies $\langle \delta^2 \rangle \sim t \cdot T^{\alpha - 1}$, where the angular brackets denote a further ensemble average. In contrast, the MSD is sub-linear in $t$, which makes CTRW non-ergodic -- the time-average and ensemble-average differ. In particular, the dependence of the TAMSD on $T$ points to the ageing nature of CTRW \cite{Metzler2014}.

A key feature of sub-diffusive CTRW is the randomness of its TAMSD. Since $P_t$ is scale free, the longest waiting times each individual trajectory encounters vary significantly, as do the amplitudes of the individual TAMSDs. To quantify this, we define the amplitude scatter, $\xi = \delta^2 / \langle \delta^2 \rangle$. For ergodic processes (e.g. $\alpha = 1$) its PDF is $P(\xi) = \delta(\xi - 1)$ for sufficiently long trajectory times. But within CTRW this PDF broadens as $\alpha$ decreases. Defining the ergodicity breaking (EB) parameter, $\EB = \langle \xi^2 \rangle - \langle \xi \rangle^2$, it can be derived analytically for CTRW processes as a monotonically decreasing function of $\alpha$.

Another cause for sub-diffusion is walking in a fractal-like environment \cite{Gefen1983, Pandey1984}. Such environment is characterised by a network of narrow passages and dead ends at different length scales, which hinder the walk. Unlike CTRW, this process is stationary and therefore ergodic. The TAMSD, like the MSD, is sub-linear in $t$, independent of $T$ and its $\EB$ parameter vanishes.

Using CTRW to model diffusion in confined geometries, such as porous media formed by either sintered or unconsolidated granular materials, is very attractive \cite{Berkowitz2006, Bijeljic2006, Wong2004, DeAnna2013} because it alleviates the need to simulate directly the dynamics of particles within the pore space, reducing significantly the computational burden. In addition, it alleviates finite-size errors due to finite samples. This practice is based on the common assumption that $P_l$, $P_t$ and $P_n$ alone control the random walk's universality class. The common procedure is to find first the forms of these distributions in a specific medium, using either small simulations or analytic derivation under some assumptions, and then use these to carry out a many-step CTRW in free space. It is then presumed that the CTRW yields the same universality class as the diffusion in the confined geometry.

The first aim of this paper is to demonstrate that this does not apply when the size of the diffusing particle is comparable to throat sizes. We do so by analysing trajectories of individual particles diffusing in a porous sample and show statistical deviations from CTRW predictions. We also compare these simulations with an equivalent CTRW model. We show that the effective change in the medium's connectivity with varying particle size affects directly the nature and universality class of the diffusion process. We conclude that the sub-diffusion is the result of CTRW on a percolation cluster. Indeed, a combination of underlying mechanisms, leading to sub-diffusion, has also been observed in \cite{Tabei2013, Weigel2011, Jeon2011, Yamamoto2014}.

The second aim of the paper is to propose a method to correct for the topological effect, which makes it possible to still use CTRW, with its advantages, to model diffusion of any finite size particle in confined geometries. To maximise the range of validity of our results (see discussion below), we consider very high porosity porous media. These correspond to marginally rigid assemblies of frictional particles, whose mean coordination number is four \cite{Blumenfeld2015}. The least confined of these are model systems whose each particle has exactly four contacts.

The structure of this paper is the following. In section \ref{sec:sample} we describe the simulated porous samples. In section \ref{sec:diffusion_in_sample} we describe the diffusion process, and discuss the effects of particle size. We perform statistical analysis of the particle trajectories and show disagreements with some predictions of the CTRW model. In section \ref{sec:diffusion_in_free_space} we describe the equivalent CTRW simulations and show that they yield different behaviour in spite of having the same step-length and waiting-time distributions. We propose an explanation for this discrepancy. In section \ref{sec:memory} we propose a modification to the conventional CTRW model to alleviate this problem, making it more suitable for modelling diffusion of finite size particles in confined geometries. We conclude in section \ref{sec:conclusions} with a discussion of the results.

\section{The porous sample} \label{sec:sample}

To simulate a three-dimensional porous granular assembly of coordination number four, we first generate an open-cell structure, using Surface Evolver as follows \cite{Brakke1992, Wang2006}. Initially, $N$ seed points are distributed randomly and uniformly within a cube, and the cube's space is Vorono\"i-tessellated to determine the cell associated with each point. A cell around a point consists of the volume of all spatial coordinates closest to it. The resulting cellular structure is then evolved with Surface Evolver to minimise the total surface area of the cell surfaces. This procedure is used commonly to model dry foams and cellular materials whose dynamics are dominated by surface tension. The result is an equilibrated foam-like structure, comprising cells, membranes, edges and vertices. A membrane is a surface shared by two neighbouring cells, an edge is the line where three membranes coincide, and a vertex is the point where four edges coincide. 

\begin{figure}[h]
        \centering
        \includegraphics[width=0.4\textwidth]{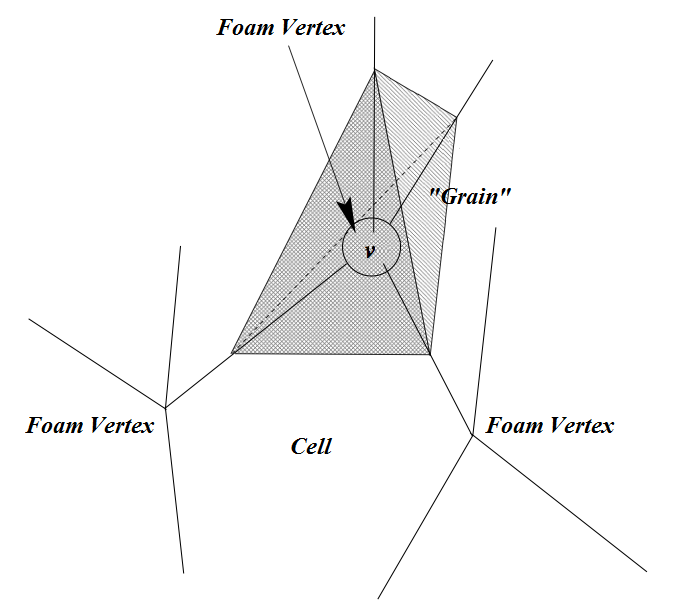}
        \caption{Pseudo-grains around the foam vertices.}
        \label{fig:pseudo_grains}
\end{figure}

Next, we construct a tetrahedron around every vertex by connecting the mid-points of the four edges emanating from it \cite{Frenkel2009}. Neighbouring tetrahedra are in contact in the sense that they share the mid-point of an edge. This construction results in a pseudo-granular structure of volume fraction $\phi = 34\%$, in which every tetrahedron represents a pseudo-grain in contact with exactly four others \cite{Blumenfeld2006} (see fig. \ref{fig:pseudo_grains}). Since neighbouring pseudo-grains share the mid-point of the edge between them, the tetrahedra structure is topologically homeomorphic to the original structure. The void space surrounded by the pseudo-grains is still cellular, but a cell surface now consists of triangular facets -- the faces of the pseudo-grains surrounding it -- and throats -- the skewed polygons remaining of the original cell membranes. The membranes over the throats are disregarded, resulting in an open-cell porous structure, in which the throats are the openings between neighbouring cells. 
The pseudo-grains volumes are smaller than those of real convex grains, which curve out into the cells of this structure. This increases the pore volume and forms a limiting case, which establishes the validity of our results for any porous medium, as will be discussed in the concluding section.

\section{Diffusion in the porous sample} \label{sec:diffusion_in_sample}

We model the diffuser as a sphere of radius $r$, measured in units of the average effective throat radius, $r_0$. We start by considering particles that are considerably smaller than the smallest throat in the structure. The particle cannot enter the tetrahedral pseudo-grains, but only move from cell $c$ to cell $c'$ through their shared throat. The simulation progresses by moving the particle from one cell, $c$, to a neighbouring cell, $c'$. Each such an event is a step, $\vec{l}_{c,c'}$, namely a vector extending between the centres of these cells. We define a waiting time, $t_c$, which is the number of time steps spent in cell $c$ before a jump occurs. Inside a cell, the particle is assumed to undergo Brownian motion and $t_c$ is proportional to (i) the square of the effective cell radius, $R_c \equiv \left(\frac{3v_c}{4\pi}\right)^{1/3}$, where $v_c$ is the cell volume, and (ii) the inverse of the fraction of the cell's open surface through which the particle can pass to neighbouring cells, $S_c\equiv \frac{\tarea}{\tarea + \farea}$. This is since the particle, on average, has to make $1/S_c$ journeys of length $R_c$ until it goes through a throat rather than hits a facet of a pseudo-grain. The effective area of a throat is the area through which a particle of radius $r$ can pass and $\tarea$ ($\farea$) is the total area of throats (facets) that make the surface of the cell. $\tarea$ is then the sum of the effective throat areas accessible for the particle to go through. Thus, the waiting time within a cell is
\begin{align} \label{eq:waiting_time}
t_c \equiv \frac{R_c^2}{2 d S_c} = \frac{1}{2d} \left( \frac{3v_c}{4\pi} \right)^{2/3} \left(1 + \frac{\farea}{\tarea} \right) \eqcomma
\end{align}
where $d$ is the local diffusion coefficient, which, using the Stokes-Einstein relation, is inversely proportional to the particle radius, $d = (r_0 / r) d_0$. 

The probability to exit cell $c$ into $c'$, $P(c' \mid c)$, is proportional to the area of the throat between them. To reduce finite-size effects, we wrap the sample around with periodic boundary conditions and let the particle travel larger distances by re-entering the sample. This means that the same cell may occur at different locations along the random walk. To avoid distorting the statistics by using the same cell too many times, we stop the process once a cell has occurred at more than five different locations.

We emphasise that we do not simulate the diffusion within the cell -- each step in our simulation corresponds to a transition of the particle from one cell to another, and the waiting time associated with the step is calculated from the cell properties. This process constitutes a random walk on a graph, whose nodes are the cell centres. After waiting for a period of time $t_c$ in cell $c$, determined by eq. (\ref{eq:waiting_time}), the particle makes a step to a neighbour cell, according to $P(c' \mid c)$.

Before continuing, it is instructive to put the problem in thermodynamic context. The cell can be regarded as a potential well of height $\Delta E$, and the probability to escape from it is $P(t) = P_0 e^{-t / \tau}$. We can then use Kramer's escape rate formula,
\begin{align} \label{eq:kramer}
\tau = \frac{2 \pi k T}{d \sqrt{U"(a)U"(b)}} e^{\Delta E / kT} \eqcomma
\end{align}
where $k$ is the Boltzmann constant, $T$ is the temperature and $U"(a)$ and $U"(b)$ are the second derivative of the potential at the bottom and top of the well, respectively. Interpreting $t_c$ as the half-life of the particle in the cell, $t_c = \tau \ln{2}$, we can combine eq. (\ref{eq:waiting_time}) and (\ref{eq:kramer}) to get:
\begin{align}
\frac{2 \ln(2) \pi k T}{\sqrt{U"(a)U"(b)}} e^{\Delta E / kT} &= \frac{1}{2 S_c(r)} \left( \frac{3v_c}{4\pi} \right)^{2/3} .
\end{align}
Assuming that all cells have the same effective potential $U$, we get:
\begin{align}
\frac{\Delta E}{kT} = {\rm Const.} - \ln{T} + \ln{ \frac{v_c^{2 / 3}}{S_c(r)} } \eqstop
\end{align}
This equation establishes the height of the effective barrier in terms of the cell volume and the fraction of its surface through which the particle can escape.

Note that the particle's mean free path within a cell is assumed to be well smaller than the cell size, regardless of the particle size, and therefore that Knudsen diffusion \cite{Knudsen1909, Clausing1930} need not be considered. However, even if this assumption is not borne out, this would only modify the coefficient $d$ in eq. (\ref{eq:waiting_time}), which is arbitrary anyway in our simulations. Also note that the above assumption, $P(t) \sim e^{-t/\tau}$, means that typical escape times do not deviate much from the mean or half-life time. This justifies our choice of taking $t_c$ as a representative.

\begin{strip}
\begin{figure}[H]
        \centering
        \subfloat[]{\includegraphics[clip,width=0.25\textwidth]{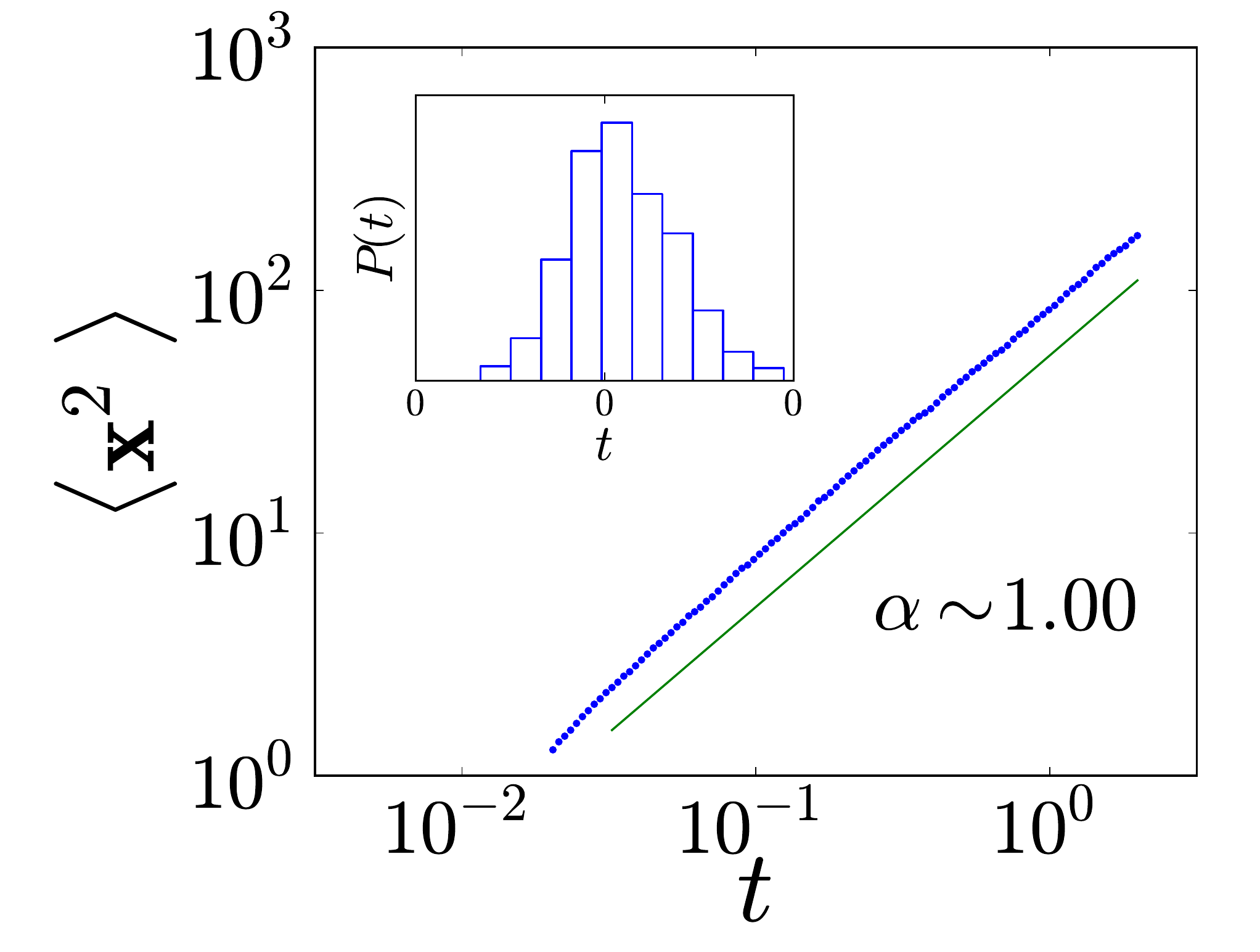} \label{fig:small_particle_in_sample}}
        \subfloat[]{\includegraphics[clip,width=0.25\textwidth]{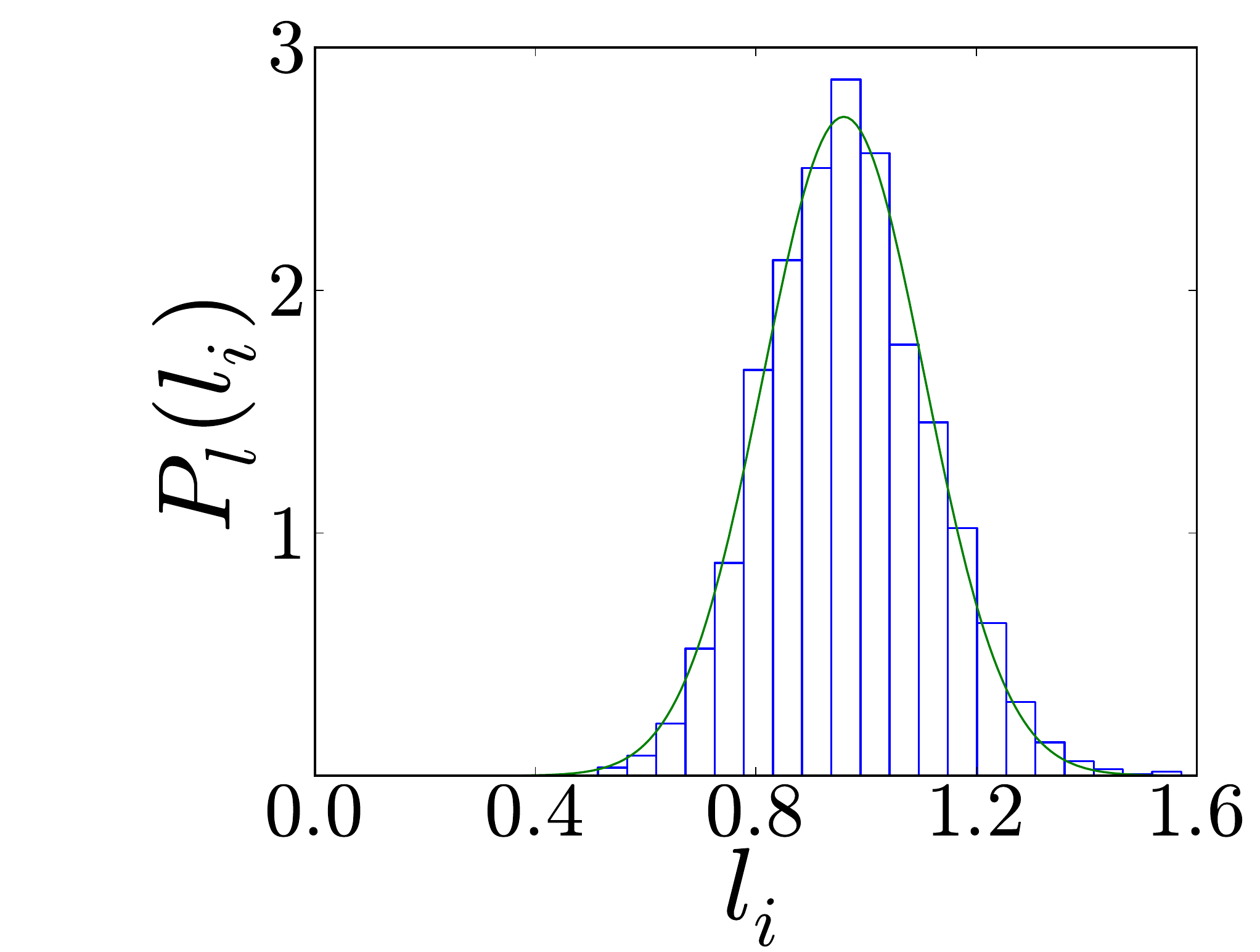} \label{fig:step_length_dist}}
        \subfloat[]{\includegraphics[clip,width=0.25\textwidth]{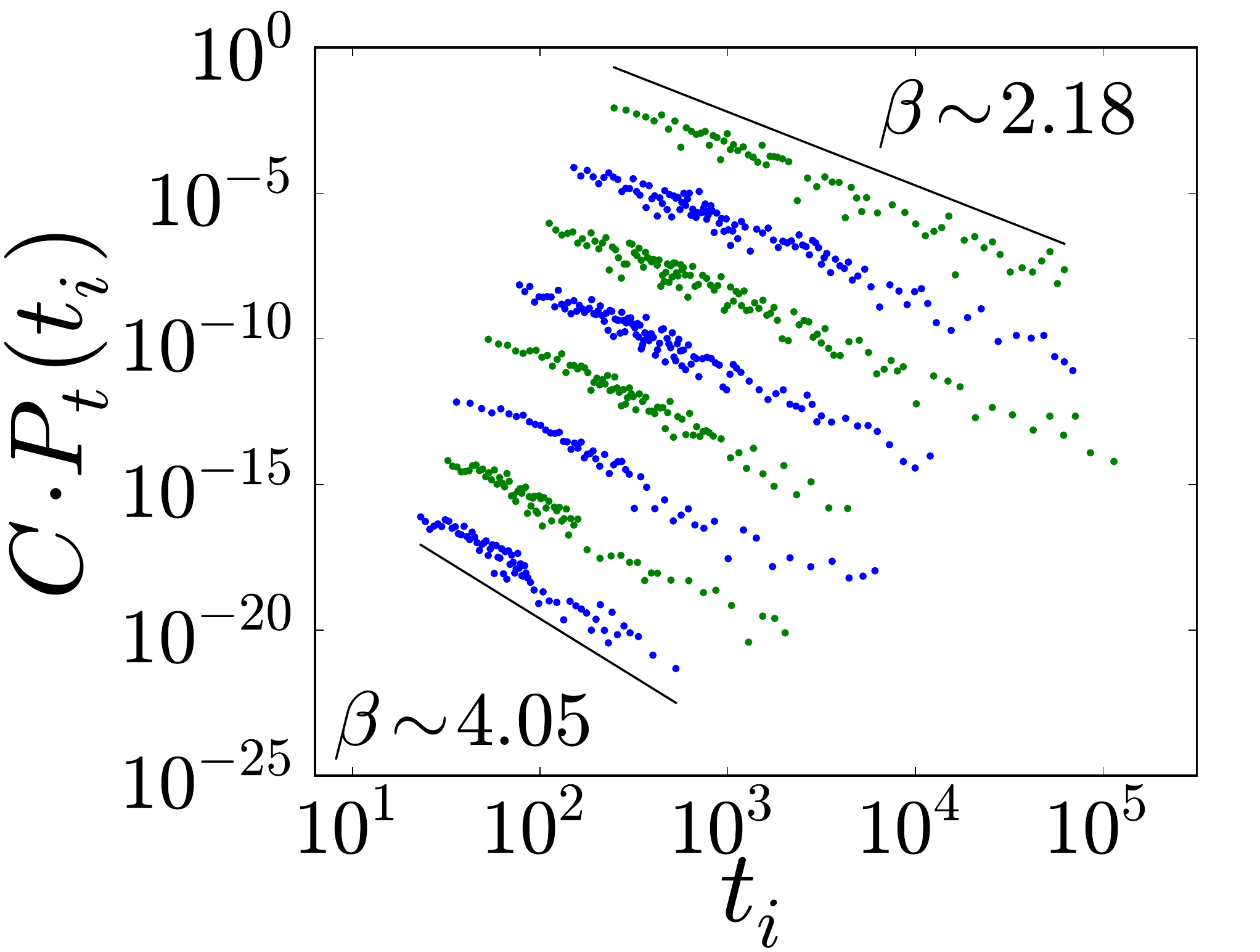} \label{fig:large_particles_waiting_time}}
        \subfloat[]{\includegraphics[clip,width=0.25\textwidth]{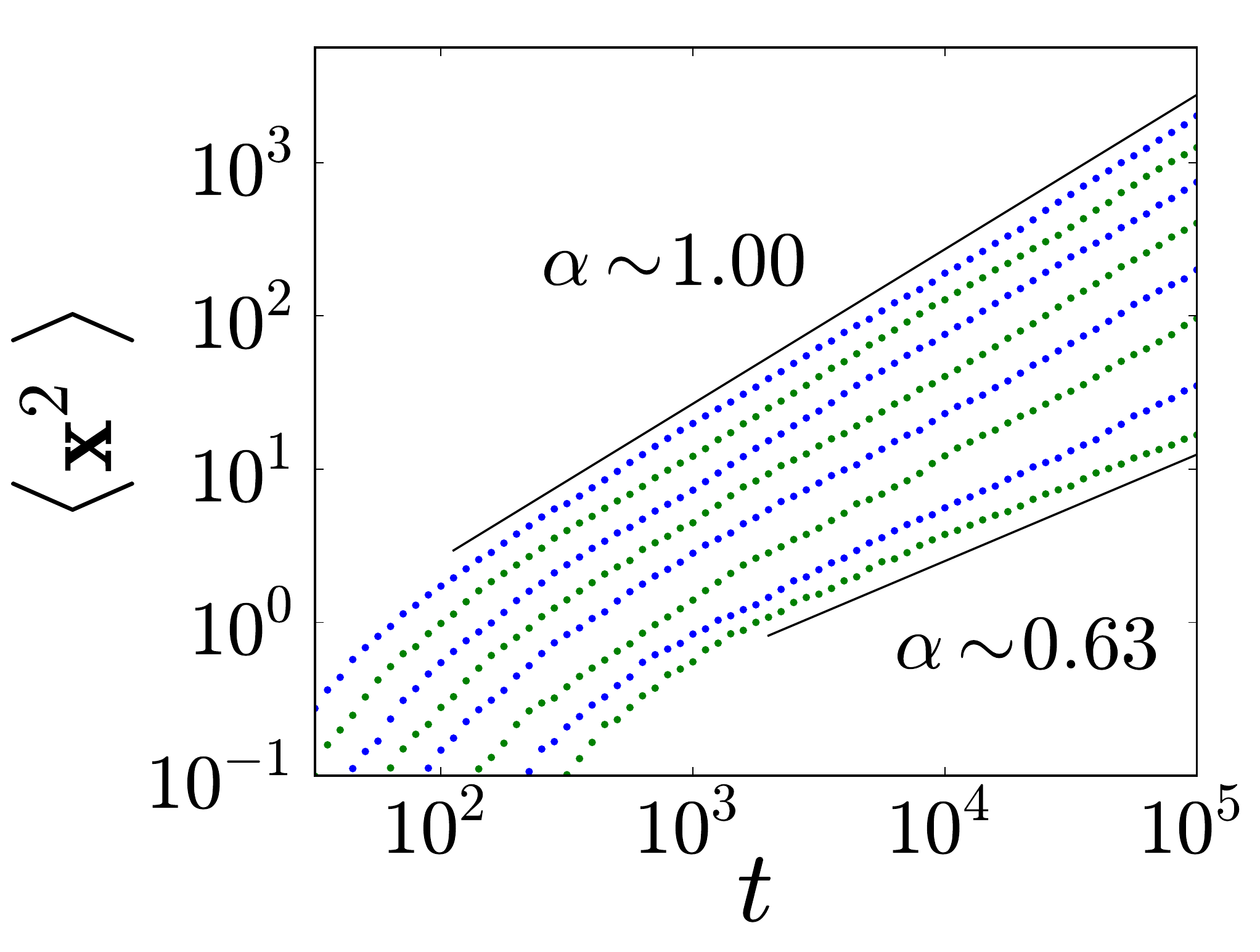} \label{fig:large_particles_msd}} \\
        \subfloat[]{\includegraphics[clip,width=0.25\textwidth]{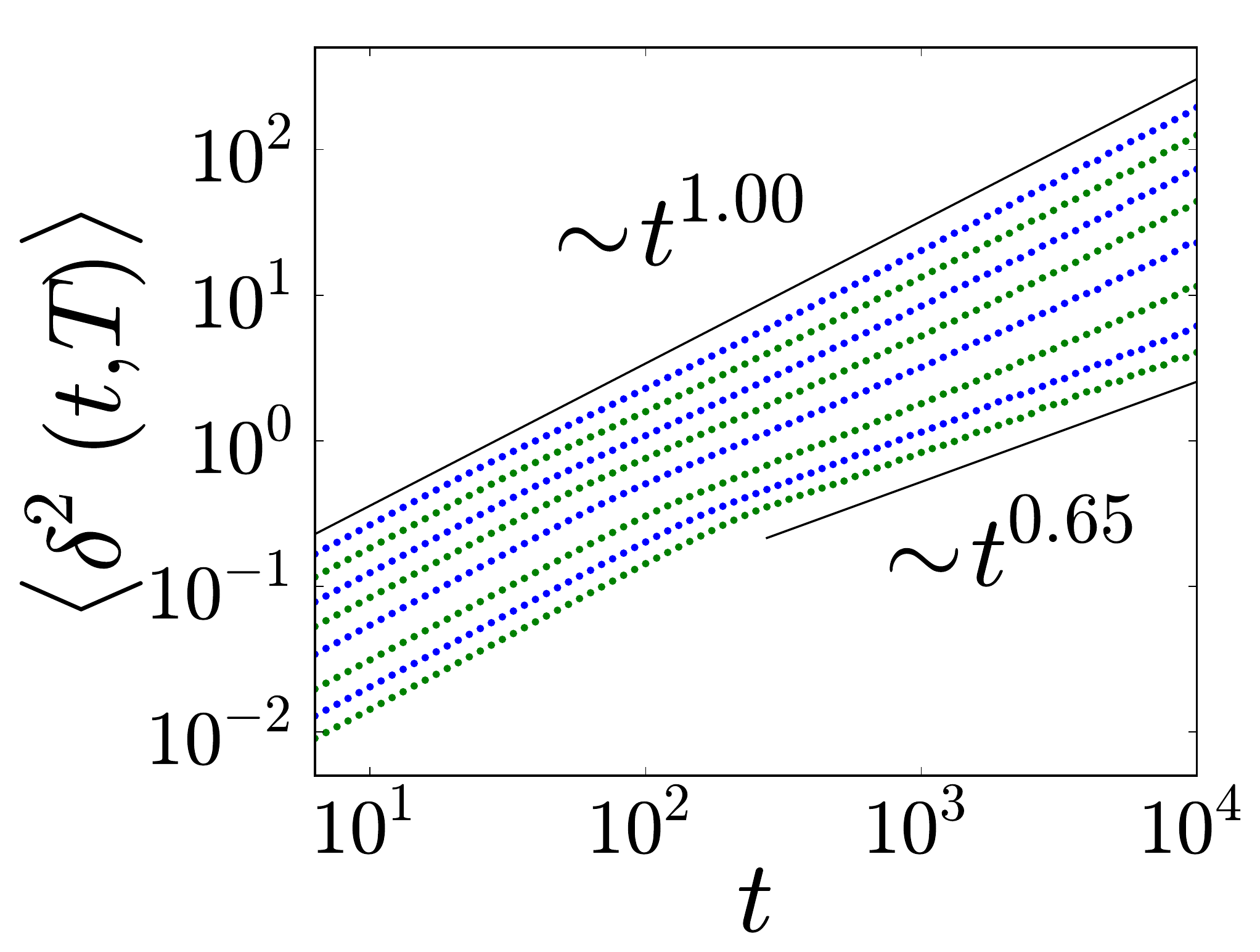} \label{fig:tamsd_vs_delta}}
        \subfloat[]{\includegraphics[clip,width=0.25\textwidth]{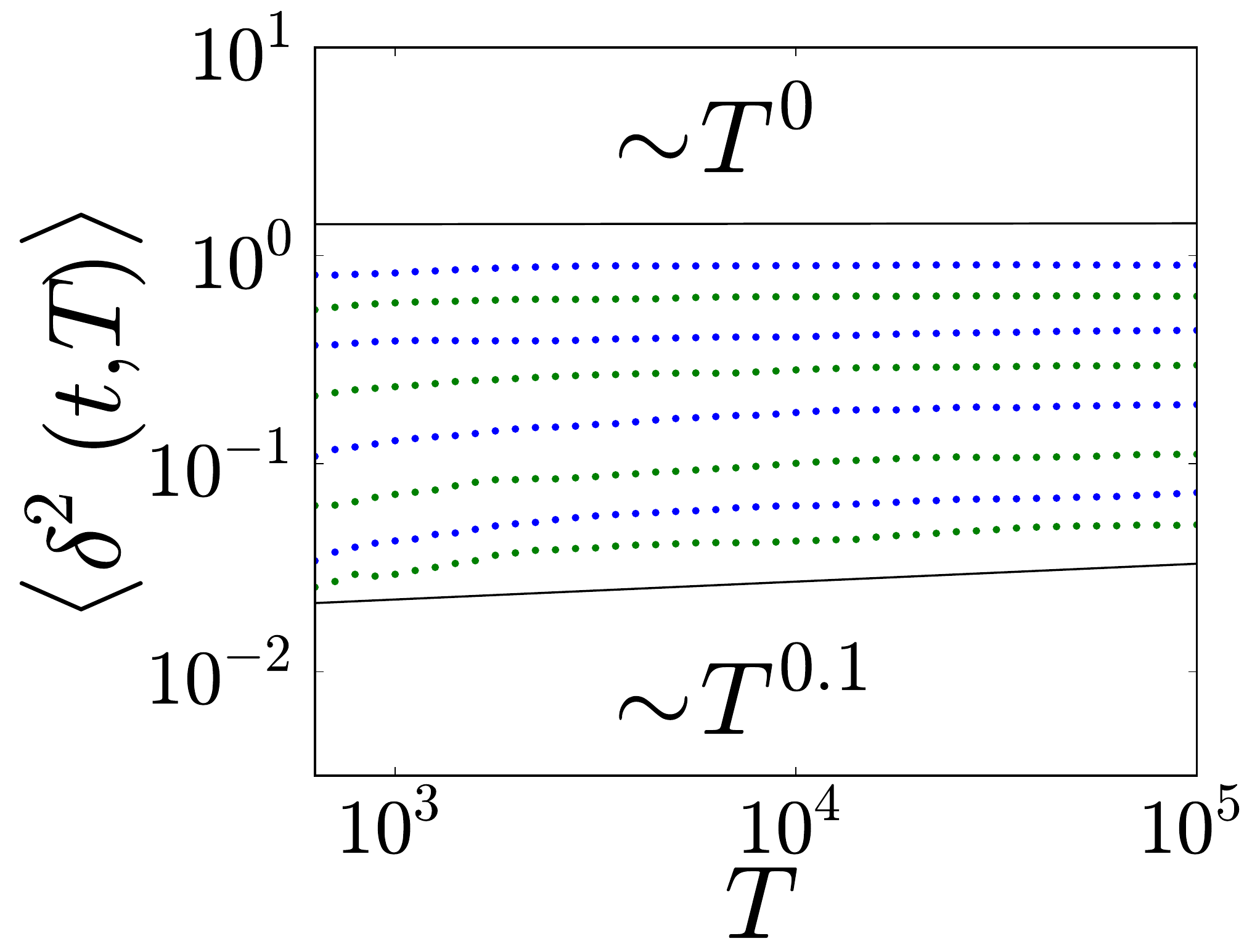} \label{fig:tamsd_vs_time}}
        \subfloat[]{\includegraphics[clip,width=0.25\textwidth]{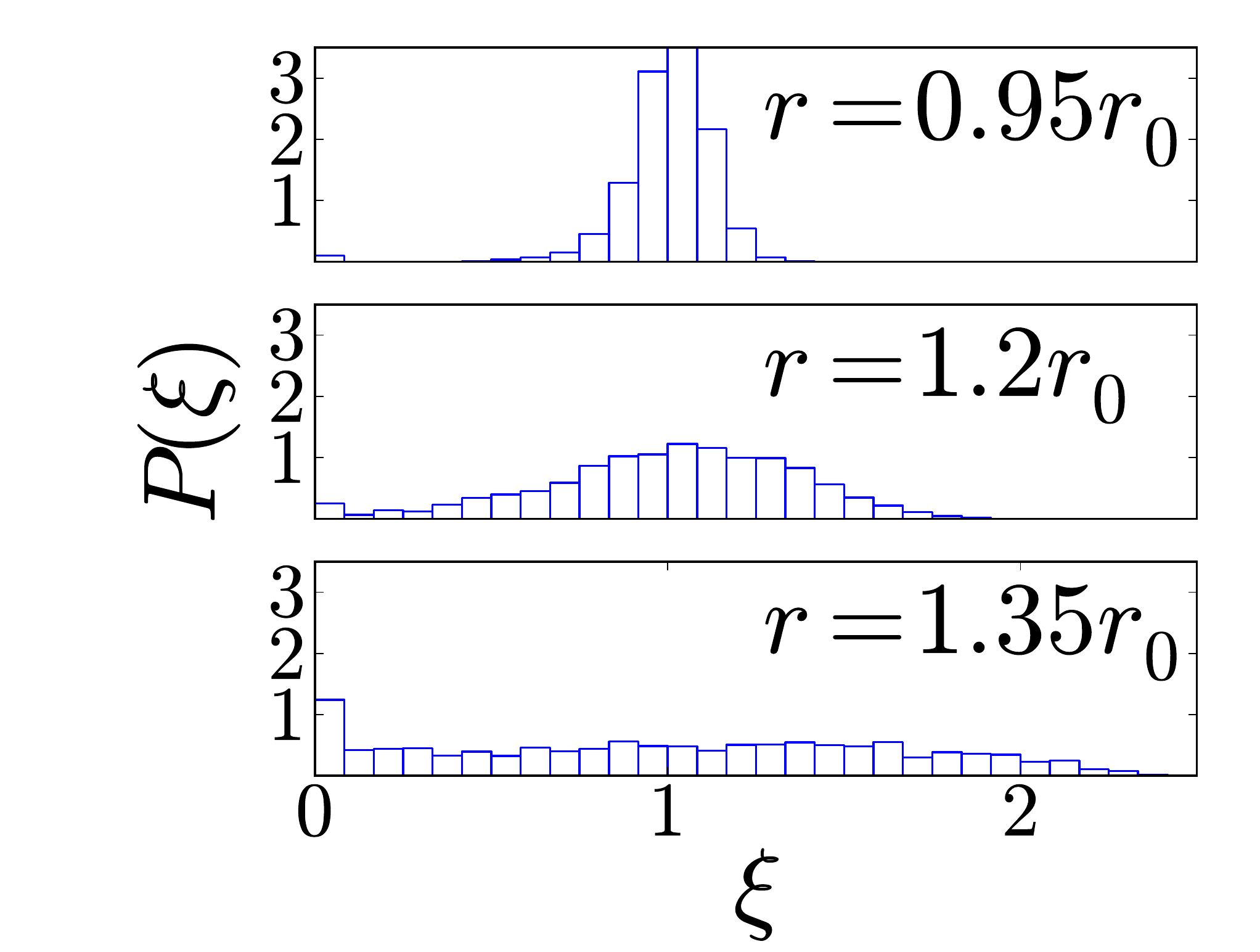} \label{fig:normalised_tamsd}}
        \subfloat[]{\includegraphics[clip,width=0.25\textwidth]{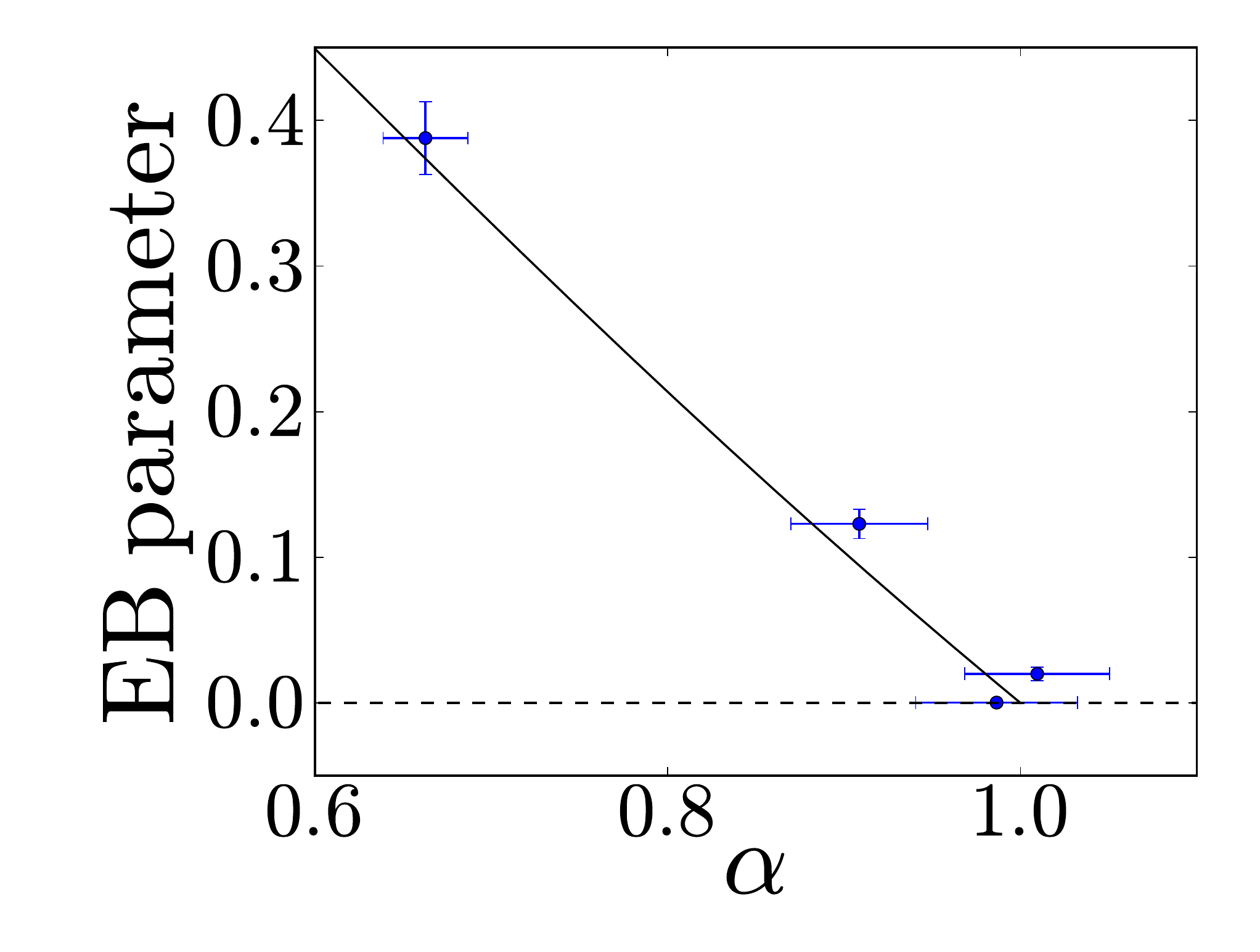} \label{fig:ergodicity_breaking}}
\end{figure}
\captionof{figure}{Statistical analysis of diffusion in the porous sample. All averages are over 1000 walks. (a) MSD vs. time for a small particle ($r = r_0 / 100$), for which $\alpha \sim 1$, indicating normal diffusion. Inset: overall waiting-time distribution of the walks. (b) Step-length distribution, $P_l(l_i)$, calculated directly from the sample (see text), with a normal fit. $P_l$ is nearly constant for all particle sizes, as the cells positions are fixed. $P_l$ is narrow, and thus does not affect the universality class of the walk. (c, d) Waiting-time distributions (left) and MSD vs. time (right) for large particles ($r/r_0 = 1, ..., 1.35$). Both $\alpha$ and $\beta$ decrease with increasing $r$. All power-law relations hold for at least 1.5 orders of magnitude. (e, f) Ensemble-averaged TAMSD, $\langle \delta^2 \rangle$, vs. $t$ (left) and vs. $T$ (right) for the same large particles. For $t > 10^{2.5}$, $\langle \delta^2 \rangle$ becomes sub-linear in $t$, like the MSD, and is nearly independent of $T$. (g) PDF of the amplitude scatter, $\xi$, from 2000 individual walks of three different particle sizes. The distribution broadens dramatically with particle size, which indicates the walk is non-ergodic. (h) The EB parameter for the same three particle sizes, as well as for $r = r_0 / 100$, vs. the anomaly parameter, $\alpha$. The solid line is the expected relation for CTRW, which goes through all points. Error bars denote $2 \sigma$ ranges, and were established by the variance of 10 independent measurements.}
\end{strip}

For each walk we calculate the particle's position at time $t$, relative to the origin, $\vec{x}(t) = \sum_{n=1}^{N(t)} \vec{l}_n$, where $N(t)$ is the number of steps made before time $t$. We then calculate the MSD, $\MSD$, as a function of $t$, where the angular brackets denote average over 1000 walks. Figure \ref{fig:small_particle_in_sample} shows $\MSD$ vs. $t$ for a particle of size $r = r_0 / 100$. The linear relation indicates normal diffusion, with a diffusion coefficient of $D = (0.65 \pm 0.03)d$ \ \ ($d = 100 d_0$). The inset figure shows a narrowly bounded $P_t$.

\paragraph{Large particles} We next consider particles of sizes comparable to $r_0$. Such particles diffuse differently due to two effects. One is delay and trapping inside cells. The larger $r$ is the lower the probability of passing through any particular throat, since the effective area, which the particle can pass through, is smaller. This reduces the overall probability to exit a cell, increasing the waiting times spent inside cells. As a result, while the waiting times are narrowly bounded for a small particle, $P_t(t_i)$ develops a power-law tail for sufficiently large particles, $P_t(t_i > t^{(0)}) \sim t_i^{-\beta}$, with $\beta$ a function of $r$. The second effect is that the topology changes with particle size; as it increases, the probability of passing through some throats vanishes identically, changing the system's connectivity for this particle.

\begin{figure}[h]
        \centering
        \includegraphics[width=0.5\textwidth]{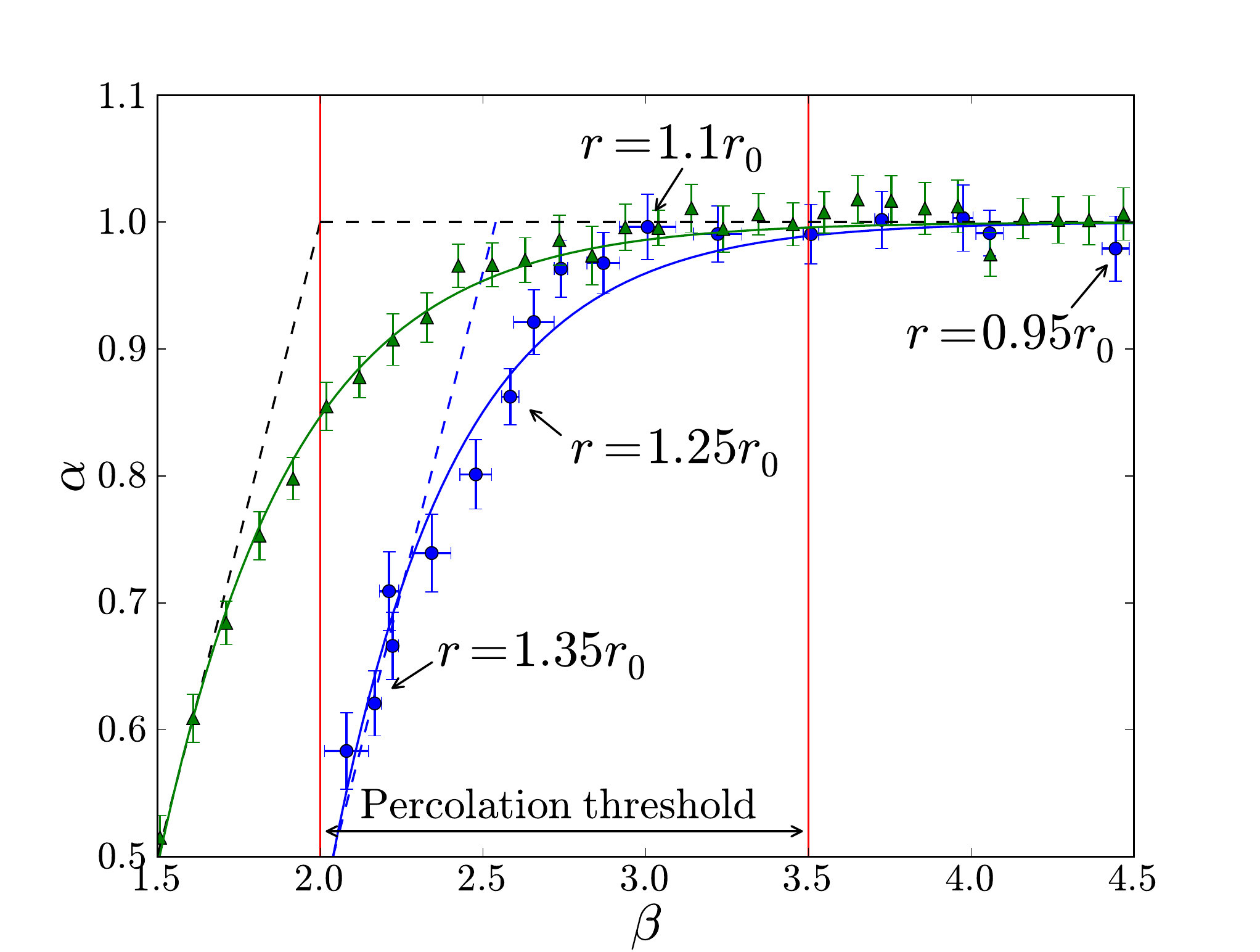}
        \caption{Two sets of simulations -- diffusion in a porous sample (blue dots) and CTRW in unconfined space (green triangles). Each simulation (dot) represents a particular particle size, and is described by the power-law of the waiting-time distribution, $\beta$, and the anomaly parameter, $\alpha$. Small (large) particles appear at the top right (bottom left) corner of the graph -- they experience a narrow (wide) waiting-time distribution and undergo normal (sub-) diffusion. The blue and green lines are fits of the form $\alpha = 1 - \frac{1}{2} \exp \{ - (\beta - \beta_t - \frac{1}{2}) / \tau \}$, with ($\beta_t^{\rm (sample)} = 2.53$, $\tau=0.42$) and ($\beta_t^{\rm (CTRW)} = 2.01$, $\tau=0.40$), respectively. The theoretical CTRW prediction of a universality class transition at $\beta=2$ is denoted by the black dashed line. The red lines mark the range of $\beta$'s that corresponds to the percolation threshold of the sample. This range matches the universality-class transition for confined diffusion. Error bars denote $2 \sigma$ ranges, and were established by the variance of 20 independent measurements.}
        \label{fig:alpha_vs_beta}
\end{figure}

Fig. \ref{fig:large_particles_waiting_time} shows $P_t$ for a few large particles. We see that $\beta$ decreases with $r$, corresponding to longer waiting times. Fig. \ref{fig:large_particles_msd} shows the MSD vs. time for the same particles. We see that beyond a certain particle size ($r \sim 1.2r_0$) $\alpha$ starts to decrease -- the diffusion becomes anomalous. To quantify this relation, we choose a larger set of radii and plot $\alpha(r)$ vs. $\beta(r)$ (blue dots in fig. \ref{fig:alpha_vs_beta}). We see that for smaller particles $\beta \gg 2$ and $\alpha=1$, corresponding to normal diffusion with narrow waiting-time PDFs. As $r$ increases, $\beta$ decreases and the walks eventually become sub-diffusive with $\alpha < 1$. We measure a transition at $\beta_t^{\rm (sample)} = 2.53 \pm 0.03$.

A short comment on scaling windows is due -- $\alpha$ is evaluated along $t \in (10^{3.5}, 10^5)$ (see fig. \ref{fig:large_particles_msd}). At much longer times, the diffusion is normal for all particle sizes. This is merely a consequence of the finite system size -- there are no cells with longer waiting times than $\sim \! 10^5$.

To investigate the diffusion process further, we calculate the TAMSD, $\delta^2$, and its average over 1000 walks, $\langle \delta^2 \rangle$. Fig. \ref{fig:tamsd_vs_delta} and \ref{fig:tamsd_vs_time} show $\langle \delta^2(t, T) \rangle$ vs. the time lag, $t$, and vs. the overall trajectory time, $T$, respectively. $\langle \delta^2 \rangle$ is sub-linear in $t$, deviating from the linear $t$-dependence in the CTRW model, and it is nearly independent of $T$. This behaviour is the same as for a random walk on a fractal. In contrast, the amplitude scatter, $\xi$, follows the CTRW prediction: Fig. \ref{fig:normalised_tamsd} shows the PDF of $\xi$ for three different particle sizes of order $r_0$. The dramatic broadening of the PDF as $r$ increases indicates that the process is non-ergodic and that there is ageing \cite{Metzler2014}. Fig. \ref{fig:ergodicity_breaking} shows that the corresponding EB parameters, i.e. the variance of these PDFs, follow the theoretical expectation from CTRW.

\section{Modelling the diffusion as isotropic CTRW} \label{sec:diffusion_in_free_space}

Next we show that an attempt to simulate this process with straightforward isotropic CTRW fails. This may not come as a surprise, as the statistical analysis showed some deviations from the traditional CTRW, but it is still constructive to describe the CTRW simulation to better understand its adjustments in section \ref{sec:memory}. For such a simulation we use the $r$-dependent PDFs, $P_l$ and $P_t$, that the particle experiences while diffusing in the confined structure. Recall that these PDFs refer to the transition between cells, rather than the movement within a cell. One way to obtain these PDFs is to measure them empirically during a diffusion process. However, more efficient is to compute them directly from the structural statistics of the porous medium, as we outline next.

A derivation of $P_l$ and $P_t$ from structural statistics should be made cautiously because a straight-forward histogram of the waiting times of all cells in the sample ignores the inherent correlation between the probability to visit a cell, $P(c)$, and the waiting time \cite{Edery2013}. Cells that are difficult to get out of (long waiting times) tend to have a lower probability of getting into and are therefore visited less frequently. In particular, some cells are completely inaccessible for particles above a certain size.

\begin{figure}[h]
        \centering
        \includegraphics[width=0.5\textwidth]{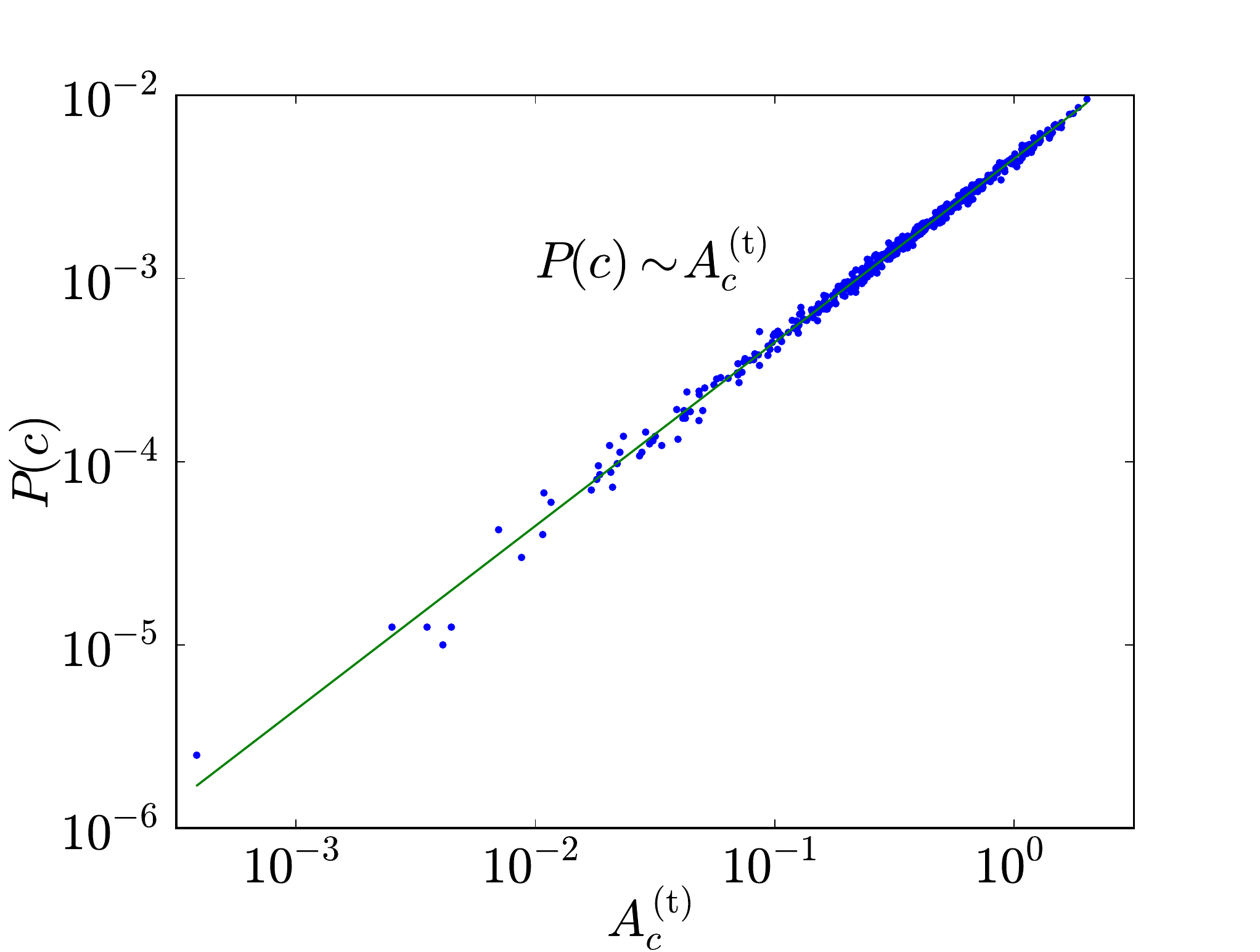}
        \caption{Visiting probability, $P(c)$, vs. the cell's total effective throat area, $\tarea$, in a diffusion process with $r = r_0$.}
        \label{fig:visits_vs_total_throat_area}
\end{figure}

To this end we use the observation that, to a very good accuracy in our diffusion process, $P(c)$ is linear in the cell's total effective throat area, $\tarea$. This observation, which is independent of the cell volume, can be seen over 3.5 orders of magnitude in fig. \ref{fig:visits_vs_total_throat_area}. Furthermore, this holds for all particle sizes, both well smaller and comparable to $r_0$. This allows us to estimate $P(c)$ for a particular particle size by using the effective $\tarea$ -- a direct structural characteristic of the medium. In principle, one expects the visiting probability to be correlated with the visiting probabilities of neighbouring cells, but fig. \ref{fig:visits_vs_total_throat_area} shows that this effect is negligible.

We can now estimate more accurately $P_t$, and, particularly, the power-law $\beta$, for every particle size, by manipulating the waiting-time histogram as follows. For every bin consisting of the waiting times of cells $\{c_1, ... , c_n\}$, we multiply the bin's height by $\sum_1^n P(c_i)$ and normalise the histogram into a PDF. This suppresses long waiting times in $P_t$.

Collecting the statistics of all possible steps in the sample, we get a PDF described well by the Gaussian $P_l(l_i) \sim \exp\{-(l_i - \bar{l})^2/2\sigma^2\}$, with $\sigma / \bar{l} = 0.15 \pm 0.02$ (see fig. \ref{fig:step_length_dist}). As $P_l$ is narrow, it does not affect the universality class of the walk. $P_l$ stays narrow, and indeed nearly constant, for all particle sizes, as the cells positions are fixed. Note that it is also possible to derive $P_l$ analytically, given the cell volume distribution and nearest neighbour volume-volume correlations. This, however, is somewhat downstream from the main thrust of this paper.

We are now able to obtain accurate step-length and waiting-time PDFs for every particle size, which could be used as input into unconstrained CTRW. Fig. \ref{fig:alpha_vs_beta} shows results of CTRW simulations (green triangles) using these PDFs. The fit to this curve (solid green line) differs from the theoretical prediction \cite{Montroll1965}, $\alpha = \beta - 1$, due to finite time and size effects. Our measured transition for CTRW is $\beta_t^{\rm (CTRW)} = 2.01 \pm 0.02$, in agreement with the theoretical prediction.

A key observation is that the CTRW exhibits a normal-to-anomalous transition at a lower values of $\beta$ than in the actual sample. Since both processes have the same step-length and waiting-time distributions, but one is performed on a graph and the other in free space, the discrepancy must stem from the connectivity of the sample, which the CTRW model cannot account for. This is supported by the statistical analysis of the diffusion in the porous sample (section \ref{sec:diffusion_in_sample}), that show behaviours typical to random walks on a fractal.

\begin{figure}[h]
        \centering
        \includegraphics[width=0.5\textwidth]{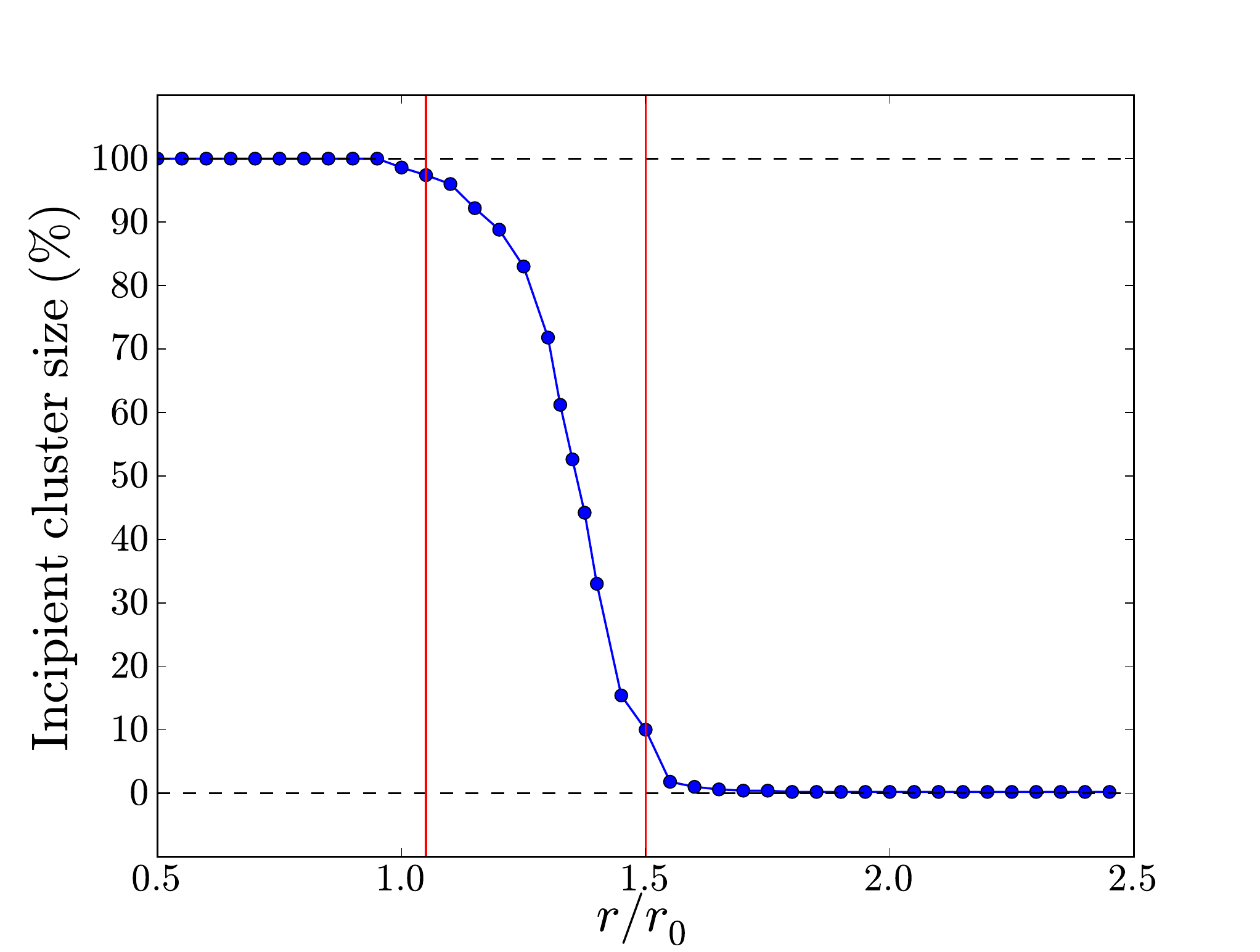}
        \caption{The percentage of cells belonging to the incipient-cluster vs. the particle radius (in units of $r_0$). The percolation range is marked in red. The corresponding $\beta$ range is marked in fig. \ref{fig:alpha_vs_beta}.}
        \label{fig:cluster_size_vs_particle_radius}
\end{figure}

As mentioned above, the size of the diffusing particle determines the effective throat sizes, and hence the connectivity of the porous sample. Moreover, above some size there is no path percolating between the sample's boundaries. Fig. \ref{fig:cluster_size_vs_particle_radius} shows the dependence of the percolating accessible volume on $r$, where a range of radii around the percolation threshold is marked by red lines. The same range is marked in fig. \ref{fig:alpha_vs_beta}. We see that the universality class transition in the sample occurs within this range. This is more evidence that the connectivity plays an important role in determining the universality class. Specifically, around the percolation threshold the incipient cluster assumes a fractal-like structure, further inducing sub-diffusion \cite{Gefen1983, Pandey1984}. We conclude that our simulations of diffusion in porous media are best described as CTRW on a percolation cluster.

\section{Anisotropic CTRW} \label{sec:memory}

\begin{figure}[h]
        \centering
        \includegraphics[width=0.5\textwidth]{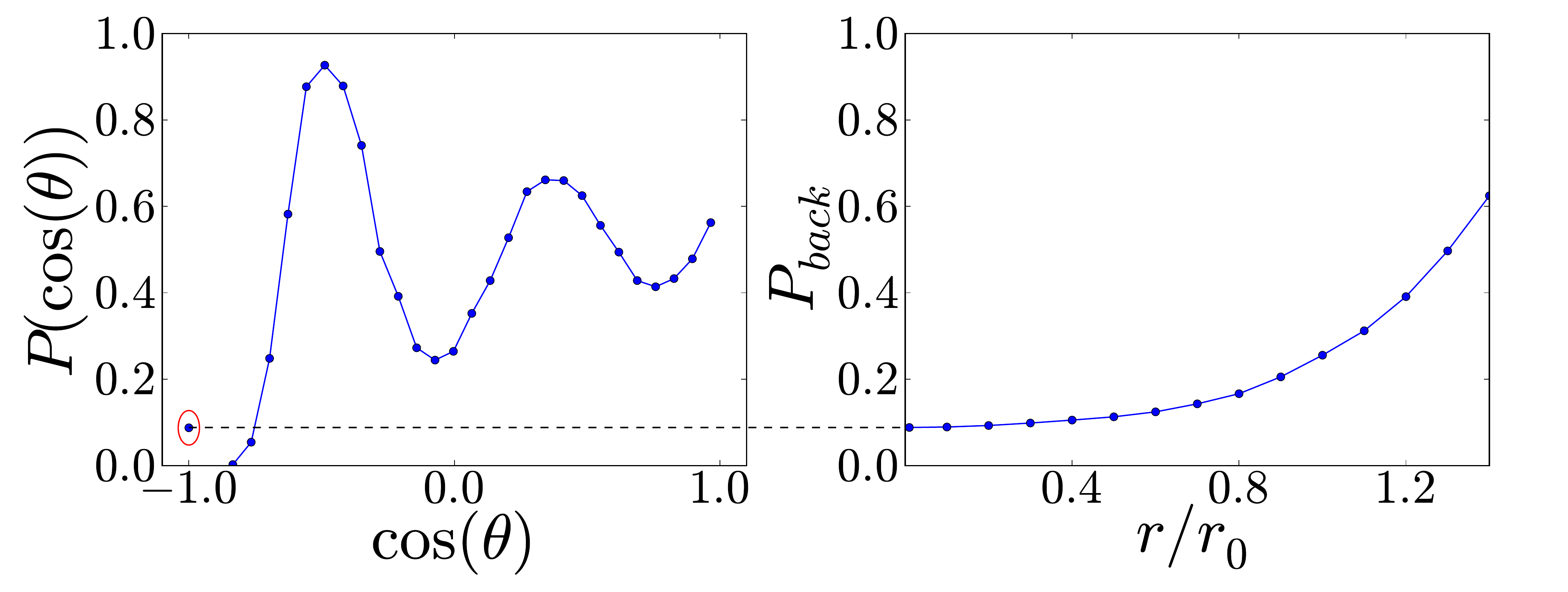}
        \caption{Left: the PDF of $\cos(\theta)$, with $\theta$ the angle between successive steps of a diffusion process in the porous sample for $r = r_0 / 100$. The singular point at $cos(\theta) = -1$ marks the finite probability of a backward step $\Pback = 0.087 \pm 0.001$. In addition, $P(-1 < \cos(\theta) < -0.88) = 0$. Right: the dependence of $\Pback$ on $r$. For large particles, backward steps dominate the walk.}
        \label{fig:direction_pdf}
\end{figure}

To retain the usefulness of the CTRW model, it would be desirable to modify it to capture the effect of connectivity. We next propose such a modification, inspired by the PDF of the angle between successive steps in the porous sample (fig. \ref{fig:direction_pdf}). There is a finite probability to make a backward step, i.e. go back through the throat the particle entered a cell, $P_{\rm back} \equiv P_\theta(\theta = \pi)$. For very small particles $\Pback = 0.087\pm 0.001$. This is in contrast to the conventional CTRW model, where the next step direction is uniformly distributed. Moreover, we see that $\Pback > 1/13.7 \approx 0.073$, which is the inverse of the average number of throats per cell, and is the expected value for $\Pback$ when all steps are equiprobable. This is because the mean size of an entrance throat is larger than the mean size of all the throats. The enhanced backward step probability can be regarded as a correlation between successively visited throats. As $r$ increases, the total available throat area decreases and $\Pback$ increases, as can be seen in the right panel of fig. \ref{fig:direction_pdf}. For very large particles, $\Pback$ dominates the walk. This can be seen as the `lowest order' effect of connectivity, which we next try to capture within CTRW.

We examine two methods to modify the CTRW model, both introducing a backward bias. In the first, we add to $P_l$ and $P_t$ a third distribution, $P_\theta$, of step direction relative to the previous step. In this method, the particle `remembers' the direction of the previous step, and a relative angle is chosen from the non-uniform distribution $P_\theta(\theta)$, e.g. the one in fig. \ref{fig:direction_pdf}. $P_\theta$ can be calculated directly from the porous structure for any particle radius $r$, similarly to $P_l$ and $P_t$ (see section \ref{sec:diffusion_in_free_space}). A random step is then made at the angle $\theta$ relative to the previous step direction. The waiting time and step length at each step are chosen as before. This method introduces a correlation between consecutive steps, which is present unavoidably in the diffusion of large particles in the sample. Testing this method by a set of simulations, we find that the universality class transition occurs at $\beta_t^{\rm (anisotropic)} = 2.1 \pm 0.03$ -- closer to the one measured in the porous sample.

The second method is more drastic: at every step of the CTRW we construct a new cell, according to the structural statistics of the porous sample. We choose the cell's volume, the number of throats, and the throats' areas from the corresponding distributions, derived from the sample. Using these and the particle size, we calculate the waiting time for the new cell. Then we choose randomly an accessible exit throat. The probability to exit through the throat is proportional to its effective area. The particle then makes a step in the direction of the exit throat. Another cell is then constructed around it. The step length is the sum of the radii of these two cells. This process is then iterated as many times as required.

A key feature of this method is that the particle `remembers' the exit direction, the last used cell and the exit throat. The latter is used as one of the throats in constructing the next cell. If this throat is chosen again then the last step is retraced. We refer to this method as DA for the two variables that the particle `remembers' -- step direction and throat area.

Within the DA process, the particle may pass back and forth several times through a large throat before it moves on and loses memory of this throat. This process correlates not only successive steps, as the previous method does, but also successive backward steps.

Using the DA model, we obtain a universality-class transition at $\beta_t^{\rm (DA)} = 2.50 \pm 0.03$, in excellent agreement with the original simulation results. Thus, this model captures much better the particle size-driven topological change.

An important feature of the DA model is that $\Pback$ is higher than in the porous sample. To understand this, consider two cells in the porous sample, connected by a large throat, and connected to the rest of the structure by smaller throats. Once the particle enters one of these cells, it is likely to move back and forth several times before it emerges. However, the smaller the throats leading to this pair of cells, the less likely is the particle to enter in the first place. This means that the occurrence frequency of such sub-structures, which increase $\Pback$, is suppressed. In contrast, once a particle passes through a relatively large throat in the DA model, then, other than this throat, an entire new cell is generated for each step. This results in a higher probability that the particle oscillates across such a throat. This feature appears to compensate for other, more complex, missing topological features, making the DA a better model for the diffusion process.

\begin{figure}[h]
        \centering
        \includegraphics[clip,width=0.5\textwidth]{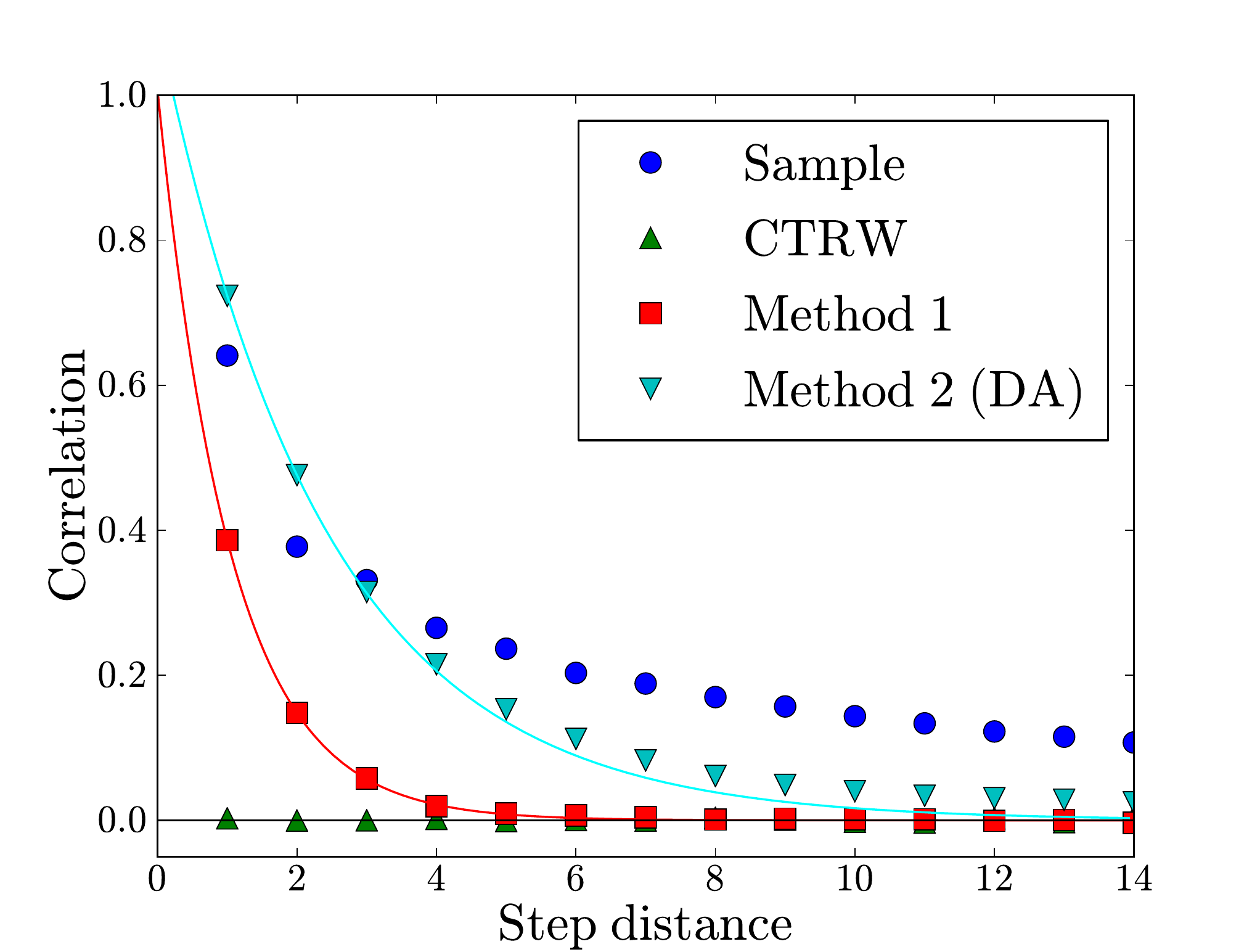}
        \caption{Step-length correlation function for the four types of simulations discussed in the paper, all with $r = 1.2r_0$. Solid lines are decaying exponentials. The correlation function of the first adjustment to CTRW decays exponentially, and the second adjustment (DA) -- more slowly.}
        \label{fig:step_length_corr}
\end{figure}

As a further investigation of the proposed methods, we present the step-length correlation function for the different models, all using $r = 1.2r_0$ (fig. \ref{fig:step_length_corr}). The correlation in the sample is mainly due to the fact that each two consecutive steps enter and exit a certain cell, $c$. If $c$ is small, then the two steps will tend to be short, and vice versa, leading to positive correlation. In addition, a high $\Pback$ means that many consecutive steps are of exactly the same length. This further increases the correlation for larger particles. As expected, the traditional CTRW exhibits no correlations. Our first adjustment to CTRW, introducing the possibility of a backward step, adds correlation. However, since this is only a one-step correlation it decays exponentially. The increase in correlation within the DA process is because of the increased probability for a long sequence of backward steps, as discussed above. As a result, the DA correlation function does not decay exponentially, agreeing better with the diffusion simulation in the sample. The correlation of the DA at a step distance of one is higher than that in the simulated diffusion process because its $\Pback$ is higher, yet the DA correlation decays faster with the number of steps since it lacks the more involved topological correlations.

\section{Conclusions} \label{sec:conclusions}

To conclude, we compared between two numerical models of diffusion of a finite size particle in porous media: a direct simulation of the diffusion process in a computer-generated sample, and what is commonly believed to be an equivalent CTRW. 
We first presented a method to construct the representative PDFs of the step-length and waiting-time, $P_l$ and $P_t$, given the particle size and statistical information about the structure of the porous media alone. 
Using the same particle size dependent $P_l$ and $P_t$ in both models, we analysed the transition from regular to anomalous diffusion. We showed that the the two models give different results -- while the CTRW simulations follow the theoretical prediction, up to finite-time effects, with a transition to sub-diffusion at $\beta \approx 2$, the diffusion simulations in the confined geometry exhibit a transition at $\beta \approx 2.5$. 
We established that the difference stems from changes to the effective connectivity available to the particle with increasing size. This particle size-driven change in connectivity is not taken into consideration in the CTRW model. We supported this conclusion by investigation of the time average MSD and by showing that the transition in universality class occurs at the same range of particle sizes that corresponds to the percolation transition. Our findings show unequivocally that the discrepancy between the two models is \textit{not} due to different waiting-time distributions, since using identical distributions do lead to different universality classes.

It is important to comment on the range of validity of our results. Increasing the particle size can be regarded, alternatively, as shrinking the porous structure, while keeping the particle size unchanged. Evidently, the smaller the pore space the more restricted the diffusing particle is and the larger the discrepancy between the simulated diffusion and the CTRW. Thus, by starting from a medium with a very large pore space, we established that our results hold true for a wide range of porous media with lower porosity.

Wong et al. \cite{Wong2004} studied experimentally a related process of trace particles diffusing in biological networks of entangled F-actin filaments. There, the diffusion of particles, of size comparable to the typical network mesh size, is sub-linear and $P_t$ decays algebraically. The universality class they observe is a function of the particle-to-mesh size ratio, in agreement with our results. However, they do not observe size-driven topological effects and their values of $\alpha$ and $\beta$ are in good agreement with the CTRW model. This is because of the flexibility of the gel-like network, which allows trapped particles to eventually escape by deforming the filament network, a phenomenon also modelled recently in \cite{Godec2014}. The rigidity of the  structure considered here precludes this particle escape mechanism and is the reason for this difference. A potentially related process of large particles, diffusing in rigid porous media, was studied experimentally in \cite{Skaug2015}. That study focused on hydrodynamic in-pore effects, which might be interesting to eventually include in our model.

The main advantages of the CTRW model are that it overcomes potential finite-size problems and is less demanding computationally. However, as we have demonstrated here, this is achieved at the expense of ignoring topological information about local connectivity. To preserve the advantages of CTRW, these need to be taken into consideration. To this end, we introduced two anisotropic CTRW models. One includes memory of the last step direction and a non-uniform distribution of step direction. The other, the DA model, adds memory of the area of the last throat visited, effectively correlating successive backward steps. The DA model shows a universality-class transition at $\beta_t^{\rm (DA)} = 2.50 \pm 0.03$, in good agreement with the one measured in our simulations of the diffusion process in the confined geometry of a porous medium. We conclude that the DA model is a better alternative to the traditional CTRW for modelling diffusion of finite size particles in such media. It combines the CTRW advantages, overcoming the finiteness of the sample and convenience of application, with a better capturing of topological effects.\\

\noindent {\bf Acknowledgement}: SA is grateful for support from the Alan Howard Scholarship.\\

\noindent {\bf Conflict of interest}: The authors declare that they have no conflict of interest.

\end{document}